\pdfoutput=1
\documentclass[10pt]{article}
\usepackage[T1]{fontenc}
\usepackage[utf8]{inputenc}
\usepackage{newtxtext,newtxmath}
\usepackage[letterpaper,textwidth=5in,textheight=7.7in,centering,headsep=14pt]{geometry}
\usepackage{amsmath,graphicx,booktabs,fancyhdr,xcolor}
\usepackage[font=footnotesize,labelfont=rm,labelsep=period,justification=centering]{caption}
\usepackage{cite}
\graphicspath{{figures/}}
\newcommand{\bibfont}{\small\setlength{\itemsep}{1.5pt}\setlength{\parsep}{0pt}}
\newcommand{\tbl}[2]{\caption{#1}\vspace{4pt}\centering\footnotesize\setlength{\tabcolsep}{3.5pt} #2}
\usepackage{titlesec}
\titleformat{\section}{\normalsize\bfseries}{\thesection.}{0.7em}{}
\titlespacing*{\section}{0pt}{16pt}{6pt}
\titleformat{\subsection}{\normalsize\bfseries\itshape}{\thesubsection.}{0.7em}{}
\titlespacing*{\subsection}{0pt}{12pt}{4pt}
\renewcommand{\thesubsection}{\thesection.\arabic{subsection}}
\fancypagestyle{plain}{\fancyhf{}}
\newcommand{\ms}[1]{\langle V_{#1}^2\rangle}
\usepackage[hidelinks]{hyperref}

\begin{document}
\thispagestyle{plain}
\begin{center}
{\large\bfseries Binary Classifier Wire-Resistance Attack on KLJN:\\ Bayes-Optimal Limit at a Narrow Resistor Gap\par}
\vspace{14pt}
{\footnotesize Mehmet Yildirim, Fahrettin Ay$^{*}$ and Laszlo B. Kish\par}
\vspace{4pt}
{\footnotesize\itshape Department of Electrical and Computer Engineering, Texas A\&M University,\\ College Station, TX 77841-3128, USA\par}
\vspace{2pt}
{\footnotesize eem.mehmetyildirim@tamu.edu, ayf@tamu.edu, laszlokish@tamu.edu\par}
\end{center}
\vspace{10pt}
\begin{center}
\begin{minipage}{4.5in}
\footnotesize
Machine learning is shown not to increase Eve's successful bit-guessing probability in the binary-classifier-based wire-resistance attack on the Kirchhoff Law-Johnson Noise (KLJN) key exchange when the resistor gap is narrow. Using time-domain simulations of a non-ideal KLJN loop with $R_L = 9$ k$\Omega$, $R_H = 10$ k$\Omega$ and wire resistances from 0 to 3200 $\Omega$, we generate 3000 secure bits per wire resistance and evaluate the mean-square noise voltages at both ends over 100-sample bit exchange periods (BEPs). This probability $p$ is evaluated for the sign rule, the nearest-line rule, logistic regression, support vector machine (SVM), $k$-nearest neighbors, random forest and gradient boosting. The sign rule gives $p = 0.711$ at 50 $\Omega$ and 0.682 at 3200 $\Omega$. The nearest-line rule, logistic regression and SVM stay close to it and the other classifiers perform worse. For Gaussian noise, the likelihood ratio between the HL and LH cases depends only on the difference of the two mean-square voltages, so the sign rule is Bayes-optimal. The resulting $p_{\max}$ reproduces the measured values and predicts that about 900 to 1200 samples per bit are needed for $p = 0.95$. This difference carries about 0.19 bit per secure bit, 1.5 times the hard-decision value.
\vspace{6pt}

\end{minipage}
\end{center}
\vspace{4pt}
{\renewcommand{\thefootnote}{}\footnotetext{\hspace{-1.8em}$^{*}$Corresponding author.}}

\section{Introduction}
\begin{figure}[htbp]
\centering
\includegraphics[width=4.6in]{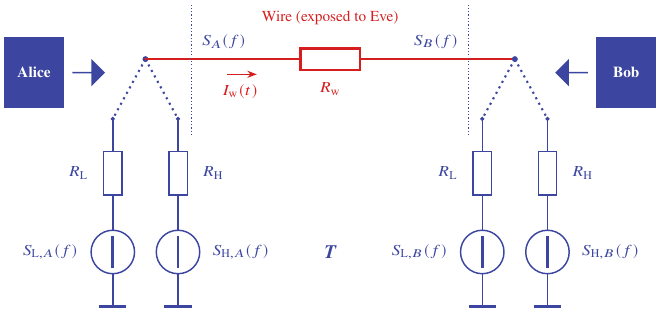}
\caption{The KLJN scheme with non-zero wire resistance $R_w$. $T$ denotes the common effective temperature. The four independent noise-voltage generators represent the thermal noise of the resistors and are typically implemented using active generators to emulate an elevated effective temperature. $S_{L,A} = S_{L,B}$ and $S_{H,A} = S_{H,B}$ are their noise spectra, and $I_w$ is the current in the wire line.}
\label{fig:scheme}
\end{figure}

Information-theoretic (unconditional) security \cite{r1,r2,r3,r4} aims to guarantee that a cryptographic system remains secure even against an adversary with unlimited computational power, so that the observed signals reveal no useful information about the secret data. In any such scheme, the security of the overall communication cannot exceed the security of its key exchange protocol. So far, only two hardware-based key exchange mechanisms have been proposed that can claim information-theoretic security: quantum key distribution (QKD) \cite{r5,r6,r7,r8,r9,r10,r11,r12,r13,r14,r15,r16,r17,r18,r19,r20,r21,r22,r23,r24,r25,r26,r27,r28,r29,r30,r31,r32,r33,r34,r35,r36,r37,r38,r39,r40,r41,r42,r43} and the classical Kirchhoff Law-Johnson Noise (KLJN) key exchange \cite{r3,r4,r44,r45,r46,r47,r48,r49,r50,r51,r52,r53,r54,r55,r56,r57,r58,r59,r60,r61,r62,r63,r64,r65,r66,r67,r68,r69,r70,r71,r72,r73,r74,r75,r76,r77,r78,r79,r80,r81,r82,r83,r84,r85,r86,r87,r88,r89,r90,r91,r92,r93,r94,r95,r96,r97,r98,r99,r100,r101,r102,r103,r104,r105,r106}. The latter (Fig.~\ref{fig:scheme}) relies solely on classical statistical physics, using thermal (Johnson) noise and Kirchhoff's circuit laws, and has been extensively analyzed over the past two decades.

In the ideal KLJN scheme \cite{r44}, Alice and Bob are connected by a perfect zero-resistance wire, and each side randomly selects one of two resistors, denoted $R_L$ and $R_H$, representing the low and high bit values. During a given clock period, the combination where Alice and Bob choose different resistors (HL or LH) constitutes a secure bit situation. In this ideal limit, the mean-square noise voltage and current on the wire are identical for the HL and LH cases, and a passive eavesdropper (Eve), who can only measure voltage and current on the line, cannot distinguish which side holds which resistor. Thus, in the absence of non-idealities and active interference, the mixed resistor states are information-theoretically secure.

However, practical KLJN implementations inevitably deviate from this idealized model. Non-zero wire resistance, cable capacitance, finite temperature accuracy, transients and other imperfections break the perfect symmetry between the HL and LH states and open the door to various attacks. One of the earliest and most studied is the wire-resistance (Bergou-Scheuer-Yariv type) attack \cite{r45,r46,r47}, where the finite resistance of the cable causes small but systematic differences between the noise statistics at Alice's and Bob's ends.

More recently, a binary-classifier-based wire-resistance attack (informally called the ``Mehmet attack'' in earlier work) has been introduced \cite{r104}. In this attack, Eve records the mean-square noise voltages at both ends of the line over many secure bits and represents each bit by a point in a two-dimensional plane defined by these mean-square values. For strongly asymmetric resistors and realistic, non-zero wire resistance, the HL and LH bits form two well-separated point clouds, enabling Eve to classify the bits with success probability $p$ close to 1, without requiring sophisticated machine learning tools. It was also noted in \cite{r104} that machine learning tools could exploit this property for enhanced detection performance. In a subsequent study \cite{r106} it was shown that narrowing the gap between the resistor values $R_L$ and $R_H$ strongly affects the strength of this attack: as the low resistor value approaches the high one, the HL and LH point clouds increasingly overlap, and for $R_L = 9$ k$\Omega$ and $R_H = 10$ k$\Omega$ the measured $p$ drops close to 0.7. That study evaluated $p$ with a simple decision rule based on the sign of the difference between the measured mean-square voltages, which we call the sign rule below, and found the surprising result that increasing the wire resistance decreases the information leak of the classifier attack.

The question we ask in this paper is whether supervised machine learning classifiers, which can use the whole two-dimensional classifier plane instead of a single sign, can recover more of the secure bits than the simple sign rule when the resistor gap is narrow. Using time-domain simulations of a non-ideal KLJN loop with $R_L = 9$ k$\Omega$, $R_H = 10$ k$\Omega$ and finite wire resistance, we generate large ensembles of secure bits and evaluate the mean-square noise voltages at Alice's and Bob's ends. Each secure bit is a point in the classifier plane, straight lines are fitted to the HL and LH point clouds, and Eve's success probability $p$ is evaluated for the nearest-line rule, for the sign rule, and for five standard supervised classifiers with cross-validation. By systematically varying the wire resistance from 0 to 3200 $\Omega$, we show that the sign rule and the best-performing classifiers stay at $p \approx 0.71$ for small wire resistance, that $p$ decreases to 0.68 at 3200 $\Omega$, and that the reason is simple: the distance between the fitted HL and LH lines stays close to the scatter of the points around them at every wire resistance. We then show that, for Gaussian noise, the likelihood ratio between the HL and LH cases depends on the samples of a BEP only through the difference of the two mean-square voltages, so the sign rule of \cite{r106} is the best possible decision, and we give its success probability as a formula of the number of samples per BEP. The formula reproduces the measured $p$ and gives the BEP length at which the attack becomes efficient. These findings confirm and extend the results of \cite{r106} and show that the security gain of resistor-gap narrowing is not lost when Eve uses machine learning.

\section{Simulations and Construction of the Classifier Data}
\subsection{Simulated KLJN loop}
The KLJN loop of Fig.~\ref{fig:scheme} was simulated in the time domain with Gaussian noise generators, in the same way as in \cite{r104,r106}, with $R_L = 9$ k$\Omega$ and $R_H = 10$ k$\Omega$. The noise voltages at Alice's and Bob's sides were generated from two independent Gaussian random streams with different seeds, so that the average power flow between the two sides vanishes in the secure states, as required for the ideal KLJN behavior \cite{r106}. The simulations used Gaussian white noise with a sampling frequency of 20 kHz, a noise bandwidth of 10 kHz and an effective noise temperature of $2 \times 10^{14}$ K, as in \cite{r104,r106}. The wire resistance took the values $R_w = 0$, 50, 100, 200, 400, 800, 1600 and 3200 $\Omega$. This range covers the practical case, where the wire resistance is a few percent of the loop resistance, and extends to hypothetical cases where the wire resistance is a large fraction of it.

For each wire resistance, 1500 secure bits were generated for the HL case and 1500 for the LH case, 3000 in total, with a bit exchange period (BEP) of 100 independent samples for each bit, and the mean-square noise voltages at Alice's and Bob's ends, $x = \ms{A}$ and $y = \ms{B}$, were recorded for every bit. With 100 independent samples per BEP, the relative standard deviation of a single mean-square value is about $\sqrt{2/100} \approx 14\%$. This is the spread of the points along the diagonal of the classifier plane in the figures below.

\subsection{Fitted lines and the geometric classifier}
For each point set a straight line $y = ax + b$ was fitted by least squares, as in \cite{r104}. Separability is determined by the distance between the fitted lines relative to the scatter of the points about those lines. When the distance is much larger than the scatter, the point sets do not overlap and Eve can read every bit. When the distance is comparable to the scatter, the point sets overlap and Eve's success probability is limited. We report the scatter as the standard deviation of the perpendicular distance of the points from their line, and the separation as the perpendicular distance between the two lines at the mean value of $x$.

Eve's classifier of \cite{r104} assigns a new point to the point set whose fitted line is nearer to it. In the cross-validation described below, the fitted lines of this nearest-line rule were obtained from the training folds and tested on the held-out fold. The decision rule of \cite{r106} uses only the sign of the difference $\ms{A} - \ms{B}$: Eve guesses HL when $\ms{A} > \ms{B}$ and LH otherwise. Both rules were evaluated on the same data.

\subsection{Machine learning classifiers}
Five standard supervised classifiers \cite{r107} were trained on the two measured quantities of each bit: logistic regression \cite{r108}, a support vector machine (SVM) with a radial basis function kernel \cite{r109}, a $k$-nearest-neighbor classifier with $k = 15$ \cite{r110}, a random forest of 300 trees \cite{r111} and gradient boosting \cite{r112}. Because the two mean-square voltages of a BEP differ only by a small fraction of their common value, the two features were given to the classifiers as the sum $u = x + y$ and the difference $v = x - y$. This is an invertible linear transformation, equivalent to expressing the data along axes rotated by 45$^\circ$, up to a common scale factor, so it does not change the information available to the classifiers. It also decorrelates the two nearly identical channels, which improves the conditioning of the linear classifiers. The features were scaled to zero mean and unit variance. Within each fold, the scaler was fitted using only the training data and then applied to the held-out fold. The default settings of the scikit-learn library \cite{r107} were used otherwise. Eve's success probability $p$ was estimated by five-fold cross-validation with the random seed 42: the 3000 bits were divided into five equal parts, each part was classified by a model trained on the other four parts, and $p$ is the fraction of correctly classified bits. In this way every bit is classified exactly once by a model that has not seen it. For the sign rule, which has nothing to train, $p$ was evaluated on all 3000 bits and a 95\% Wilson confidence interval \cite{r113} was attached. For the logistic regression we also recorded the area under the receiver operating characteristic (ROC) curve (AUC), a standard measure of how well the classifier ranks the bits from most likely HL to most likely LH.

\subsection{Optimal decision rule and its success probability}
Within one BEP, the $N$ samples $\mathbf{v}_n = (V_A(t_n), V_B(t_n))^{\rm T}$ of the noise voltages at the two ends are, for Gaussian noise, independent samples of a two-dimensional Gaussian distribution with zero mean and covariance matrix $\Sigma$. Exchanging the resistors of Alice and Bob exchanges the two ends, so the covariance matrix of the LH case follows from that of the HL case by swapping the two coordinates,
\begin{equation}
\Sigma_{\rm HL} = \Sigma = \begin{pmatrix} \Sigma_{AA} & \Sigma_{AB} \\ \Sigma_{AB} & \Sigma_{BB} \end{pmatrix}, \qquad \Sigma_{\rm LH} = P\,\Sigma\,P, \qquad P = \begin{pmatrix} 0 & 1 \\ 1 & 0 \end{pmatrix},
\label{eq:cov}
\end{equation}
with $\Sigma_{AA} > \Sigma_{BB}$ in the HL case. The likelihood of the $N$ samples depends on them only through the sample covariance matrix $S$, whose elements are the three second-order sample moments of the BEP, $x = \ms{A}$, $y = \ms{B}$ and $c = \langle V_A V_B\rangle$. The logarithm of the likelihood ratio between the two cases is
\begin{equation}
\Lambda = \frac{N}{2}\left[\operatorname{tr}\!\left(\Sigma_{\rm LH}^{-1} S\right) - \operatorname{tr}\!\left(\Sigma_{\rm HL}^{-1} S\right)\right] = \frac{N}{2}\,\frac{\Sigma_{AA} - \Sigma_{BB}}{\det\Sigma}\,(x - y),
\label{eq:lr}
\end{equation}
because the swap preserves the determinant, $\det\Sigma_{\rm LH} = \det\Sigma_{\rm HL}$, so the normalization terms of the two Gaussians cancel, and $\Sigma_{\rm LH}^{-1} - \Sigma_{\rm HL}^{-1}$ is diagonal with two entries of opposite sign, so the cross term $c$ cancels as well. As Eq.~(\ref{eq:lr}) shows, the decision rule of \cite{r106}, to guess HL when $x > y$, is the likelihood ratio test, that is, the best possible decision (the Bayes-optimal decision) for equally probable bits. Every classifier that works on the noise voltages of a single BEP is bounded by it, however complicated it is, and Eve does not need to know $\Sigma$ to use it.

The success probability of this rule follows from the distribution of the difference $x - y$, which can be written as $N^{-1}\sum_n \mathbf{v}_n^{\rm T} D\, \mathbf{v}_n$ with $D = {\rm diag}(1, -1)$. The matrix $D\Sigma$ has one positive and one negative eigenvalue, $\lambda_\pm = \sqrt{\det\Sigma}\,\bigl(\kappa \pm \sqrt{1 + \kappa^2}\bigr)$, where
\begin{equation}
\kappa = \frac{\Sigma_{AA} - \Sigma_{BB}}{2\sqrt{\det\Sigma}}
\label{eq:kappa}
\end{equation}
is the asymmetry parameter of the loop. Therefore $x - y$ is the difference of two independent chi-squared random variables with $N$ degrees of freedom, weighted by $\lambda_+$ and $|\lambda_-|$, and
\begin{equation}
p_{\max}(N) = \Pr\!\left(F_{N,N} > \frac{\sqrt{1 + \kappa^2} - \kappa}{\sqrt{1 + \kappa^2} + \kappa}\right) \approx \Phi\!\left(\frac{\kappa\sqrt{N}}{\sqrt{1 + 2\kappa^2}}\right),
\label{eq:pmax}
\end{equation}
where $F_{N,N}$ is an $F$-distributed random variable with $N$ and $N$ degrees of freedom, $\Phi$ is the standard normal distribution function, and the second form is the Gaussian approximation for large $N$. Thus Eq.~(\ref{eq:pmax}) gives the maximum success probability for any BEP length from the single number $\kappa$, which depends on the resistors and on the wire resistance but not on $N$.

The covariance matrix was estimated from the simulated data as $\Sigma_{AA} = \langle x\rangle$, $\Sigma_{BB} = \langle y\rangle$ and, from ${\rm Var}(x - y) = (2/N)\,(\Sigma_{AA}^2 + \Sigma_{BB}^2 - 2\Sigma_{AB}^2)$, the cross term $\Sigma_{AB}^2$. A second estimate of $\Sigma_{AB}^2$ from ${\rm Cov}(x, y) = 2\Sigma_{AB}^2/N$ agrees with it within about 0.5\% (0.7\% at the largest wire resistance), and the number of independent samples per BEP obtained from ${\rm Var}(x) = 2\Sigma_{AA}^2/N$ is 100.5. This confirms the assumptions behind Eqs.~(\ref{eq:cov}) to (\ref{eq:pmax}) for the simulated data. The exact form of Eq.~(\ref{eq:pmax}) was also checked by a computer simulation of $2 \times 10^5$ BEPs of 100 samples drawn from the estimated covariance matrix.

\section{Results}
\subsection{The classifier plane}
\begin{figure}[htbp]
\centering\includegraphics[width=5in]{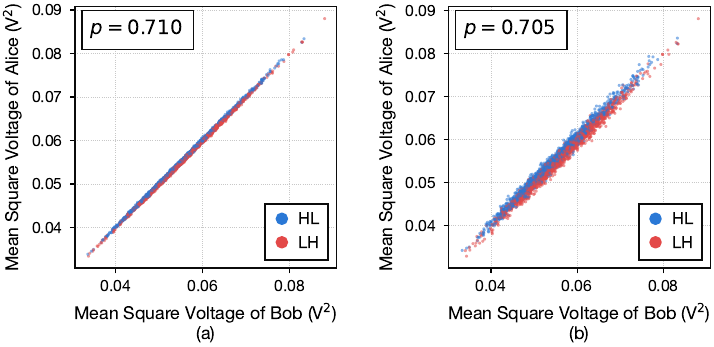}
\caption{Classifier plot of the mean-square voltages of Alice and Bob for $R_L = 9$ k$\Omega$ and $R_H = 10$ k$\Omega$ with (a) 200 $\Omega$ and (b) 800 $\Omega$ wire resistances, 1500 BEPs per point set and BEP = 100 samples. Eve's successful bit guessing probabilities are $p = 0.710$ and $p = 0.705$, respectively.}
\label{fig:cls1}
\end{figure}

\begin{figure}[htbp]
\centering\includegraphics[width=5in]{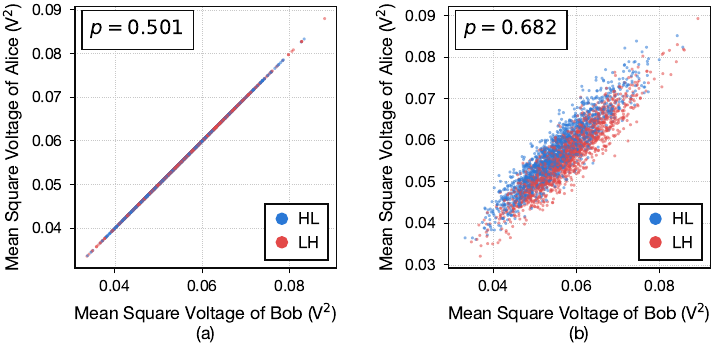}
\caption{Classifier plot of the mean-square voltages of Alice and Bob for $R_L = 9$ k$\Omega$ and $R_H = 10$ k$\Omega$ with (a) zero wire resistance, where the two mean-square voltages are identical for every BEP and Eve's successful bit guessing probability is $p = 0.501$, that is, the ideal limit of 0.5, and (b) the hypothetical case of 3200 $\Omega$ wire resistance, where $p = 0.682$.}
\label{fig:cls2}
\end{figure}

Figures~\ref{fig:cls1} and \ref{fig:cls2} show the classifier plots of the mean-square voltages of Alice and Bob for wire resistances of 200 and 800 $\Omega$ and for the two extreme cases of 0 and 3200 $\Omega$, with all 1500 BEPs of each point set. The success probability $p$ of the sign rule is shown in each panel, and the parameters of the lines fitted to the two point sets are listed in Table~\ref{tab:lines}.

At $R_w = 0$ the mean-square voltages at the two ends are identical for every BEP, both fitted lines coincide with $y = x$, and $p = 0.5$: Eve cannot extract any information, as expected. At 200 $\Omega$ the fitted slopes are 0.9971 for HL and 1.0006 for LH, that is, the two lines differ by only about 0.35\% in slope, and the two point sets lie on top of each other along the whole diagonal. At 800 $\Omega$ the slopes are 0.9845 and 0.9970. The points scatter away from the lines much more than at 200 $\Omega$ because the wire resistance decouples the noise generators of Alice and Bob, exactly as observed in \cite{r104,r106}, but the two point clouds still overlap. At 3200 $\Omega$ the slopes are 0.8961 and 0.9278 and the scatter is so large that the separation of the lines is again not visible in the point clouds. The HL and LH sets do not separate in any of the panels. This is completely different from the strongly asymmetric resistor pairs of \cite{r104,r106}, where the two sets form distinct lines without overlap.

\subsection{Overlap of the point sets}
\begin{figure}[htbp]
\centering\includegraphics[width=5in]{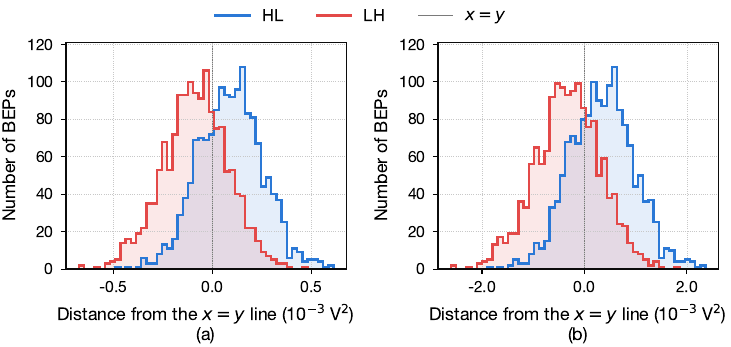}
\caption{Distribution of the perpendicular distance of the HL and LH points from the line $x = y$ of the classifier plane at (a) 200 $\Omega$ and (b) 800 $\Omega$ wire resistances, 1500 BEPs per set. Positive distance means that Alice's mean-square voltage is larger than Bob's. The two distributions overlap strongly.}
\label{fig:hist}
\end{figure}

The information that matters for Eve lies in a very thin band around the diagonal of the classifier plots. Figure~\ref{fig:hist} therefore shows the distribution of the perpendicular distance of the points from the line $x = y$ for the HL and LH sets at 200 and 800 $\Omega$. The two distributions are mirror images of each other, as expected from the symmetry of the HL and LH cases, and they are shifted in opposite directions by a small amount, but each of them extends to both sides of zero: about 29\% of the HL points lie on the LH side of the line $x = y$ and vice versa. This is the only source of Eve's errors.

Table~\ref{tab:lines} gives the numbers behind these plots. The slopes of the fitted lines move away from 1 with increasing wire resistance, with the HL slope changing faster than the LH slope, and the distance between the two lines grows from $0.045 \times 10^{-3}$ V$^2$ at 50 $\Omega$ to $2.2 \times 10^{-3}$ V$^2$ at 3200 $\Omega$. The scatter of the points around their own line grows at the same rate, from $0.04 \times 10^{-3}$ V$^2$ to $2.3 \times 10^{-3}$ V$^2$. Therefore the ratio of separation to scatter, which decides whether the point sets overlap, stays between 1.1 and 1.0 over the whole range from 50 to 1600 $\Omega$ and falls slightly below 1 at 3200 $\Omega$. A ratio of about 1 means that the two point clouds are shifted from each other by only one standard deviation of their width. This is the reason for the overlap seen in the classifier plots and for the upper limit of Eve's success probability.

\begin{table}[htbp]
\tbl{Eve's success probability $p$ for the different classifiers as a function of the wire resistance $R_w$, and the upper limit of Eq.~(4) evaluated with the covariance matrix estimated from the data. The sign rule is evaluated on all 3000 bits and is given with the half-width of its 95\% Wilson confidence interval (CI); the other classifiers are evaluated by stratified five-fold cross-validation (mean over the folds). At $R_w = 0$ the values of the machine learning classifiers, which use the sum and difference features, reflect a finite-sample offset between the two simulation runs (Sec.~3.3); with the difference alone, all of them give 0.50.}{
\begin{tabular}{@{}rcccccccc@{}}\toprule
$R_w$ ($\Omega$) & Limit & Sign rule ($\pm$CI) & Nearest line & Log. regr. & SVM & $k$-NN & Rand. forest & Grad. boost. \\ \midrule
0 & 0.500 & 0.501 $\pm$ 0.018 & 0.501 & 0.531 & 0.524 & 0.509 & 0.503 & 0.514 \\
50 & 0.708 & 0.711 $\pm$ 0.016 & 0.710 & 0.703 & 0.699 & 0.673 & 0.657 & 0.699 \\
100 & 0.708 & 0.710 $\pm$ 0.016 & 0.709 & 0.702 & 0.699 & 0.674 & 0.658 & 0.691 \\
200 & 0.707 & 0.710 $\pm$ 0.016 & 0.708 & 0.701 & 0.699 & 0.674 & 0.654 & 0.694 \\
400 & 0.705 & 0.707 $\pm$ 0.016 & 0.706 & 0.700 & 0.697 & 0.670 & 0.659 & 0.697 \\
800 & 0.701 & 0.705 $\pm$ 0.016 & 0.702 & 0.697 & 0.696 & 0.669 & 0.650 & 0.694 \\
1600 & 0.694 & 0.698 $\pm$ 0.016 & 0.691 & 0.691 & 0.689 & 0.658 & 0.650 & 0.680 \\
3200 & 0.682 & 0.682 $\pm$ 0.017 & 0.676 & 0.678 & 0.674 & 0.652 & 0.641 & 0.670 \\
\bottomrule\end{tabular}}
\label{tab:p}
\end{table}

\begin{table}[htbp]
\tbl{Parameters of the least-squares lines $y = ax + b$ fitted to the HL and LH point sets in the plane $x = \langle V_A^2\rangle$, $y = \langle V_B^2\rangle$ (all 1500 BEPs per set), the scatter of the points about their own line (one standard deviation of the perpendicular distance, HL set), the perpendicular separation between the two lines at the mean value of $x$, their ratio, and the area under the receiver operating characteristic curve (AUC) of the logistic regression. Intercepts, scatter and separation are given in units of $10^{-3}$\,V$^2$. At $R_w = 0$ both lines coincide with $y = x$ and the ratio is undefined.}{
\begin{tabular}{@{}rcccccccc@{}}\toprule
$R_w$ ($\Omega$) & $a_{\rm HL}$ & $b_{\rm HL}$ & $a_{\rm LH}$ & $b_{\rm LH}$ & Scatter & Separation & Ratio & AUC \\ \midrule
0 & 1.0000 & 0.00 & 1.0000 & 0.00 & 0.000 & 0.000 & n/a & 0.53 \\
50 & 0.9993 & 0.00 & 1.0002 & 0.02 & 0.040 & 0.045 & 1.10 & 0.78 \\
100 & 0.9986 & 0.01 & 1.0004 & 0.04 & 0.080 & 0.089 & 1.10 & 0.78 \\
200 & 0.9971 & 0.03 & 1.0006 & 0.09 & 0.160 & 0.177 & 1.09 & 0.78 \\
400 & 0.9935 & 0.11 & 1.0002 & 0.23 & 0.317 & 0.347 & 1.09 & 0.78 \\
800 & 0.9845 & 0.37 & 0.9970 & 0.63 & 0.621 & 0.669 & 1.07 & 0.77 \\
1600 & 0.9604 & 1.32 & 0.9820 & 1.85 & 1.198 & 1.245 & 1.03 & 0.76 \\
3200 & 0.8961 & 4.37 & 0.9278 & 5.51 & 2.250 & 2.171 & 0.95 & 0.75 \\
\bottomrule\end{tabular}}
\label{tab:lines}
\end{table}

\subsection{Eve's success probability}
\begin{figure}[htbp]
\centering\includegraphics[width=4.0in]{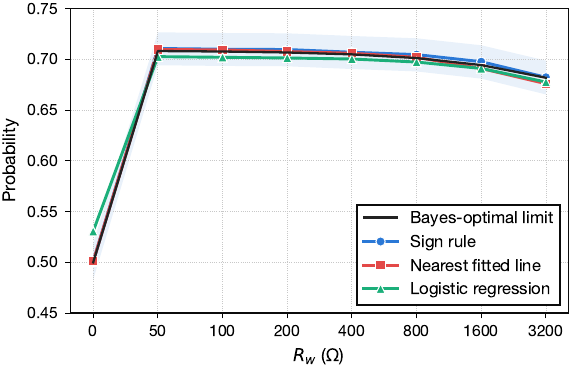}
\caption{Eve's successful bit-guessing probability $p$ versus the wire resistance $R_w$ for Alice's/Bob's resistors of 9/10 k$\Omega$: the Sign rule of \cite{r106} (band: 95\% confidence interval from 3000 bits), the Nearest fitted line of \cite{r104} and Logistic regression (both by five-fold cross-validation). The thin black line is the Bayes-optimal limit of Eq.~(\ref{eq:pmax}). No learned classifier systematically outperforms the sign rule. The measured results are consistent with the Bayes-optimal limit within finite-sample fluctuations. These results further confirm the counterintuitive nature of the behavior of $p$ versus $R_w$ in the classifier attack.}
\label{fig:p}
\end{figure}

Table~\ref{tab:p} and Fig.~\ref{fig:p} summarize Eve's success probability for all classifiers. Three observations follow.

First, for non-zero wire resistance the sign rule gives $p = 0.711$ at 50 $\Omega$, 0.710 at 100 and 200 $\Omega$, 0.707 at 400 $\Omega$, 0.705 at 800 $\Omega$, 0.698 at 1600 $\Omega$ and 0.682 at 3200 $\Omega$, with a 95\% confidence interval of about $\pm 0.016$ on each value. The values at 100, 200, 400 and 800 $\Omega$ agree with the $p = 0.709$, 0.706, 0.705 and 0.704 reported for the same resistor pair in \cite{r106} within the statistical uncertainty of the 3000 simulated bits. At $R_w = 0$ the sign rule gives $p = 0.501$, that is, the ideal limit $p = 0.5$ within the statistical uncertainty.

Second, for non-zero wire resistance, none of the machine learning classifiers systematically outperforms the sign rule. The nearest-line rule of \cite{r104} gives the same values within 0.007. Logistic regression, which can place its decision line anywhere in the plane, gives $p$ between 0.703 and 0.678, within 0.01 of the sign rule at every non-zero wire resistance, and the SVM gives $p$ within 0.012 of it. At zero wire resistance, logistic regression and the SVM give 0.531 and 0.524, above the sign rule. This is not a property of the HL and LH cases: at $R_w = 0$ the difference $v$ is zero for every bit, and the only remaining input is the sum $u$, in which the two separately simulated runs of the HL and LH cases differ by 1.5\% in their average level. This 1.5\% difference is a finite-sample offset between the independent random-number realizations of the two runs and has no physical significance: the same realizations are used at every wire resistance, and the offset is the same 1.5\% at every wire resistance, whereas a physical asymmetry between the HL and LH cases would change with the wire resistance. The classifiers learn this offset. Using $v$ alone changes the results of logistic regression, SVM, and gradient boosting by less than 0.012, and none exceeds the sign rule by more than 0.003. With $v$ alone, all five classifiers give $p = 0.50$ at zero wire resistance. In the original two-feature comparison, $k$-NN, random forest, and gradient boosting remain the poorer-performing classifiers, with $p$ between 0.64 and 0.70, and their lower performance is consistent with overfitting to finite-sample fluctuations, since they divide the plane into many small regions instead of using the single direction that carries the information. The AUC of the logistic regression, between 0.78 at 50 $\Omega$ and 0.75 at 3200 $\Omega$, shows that the size of the difference $\ms{A} - \ms{B}$ carries some extra information about how likely a guess is to be right, but this information cannot be turned into a better bit-by-bit decision, as discussed in Sec.~4.

Third, $p$ decreases with increasing wire resistance for the sign rule, and the classifiers show an overall decreasing trend. The decrease is slow between 50 and 800 $\Omega$ and becomes faster above 1600 $\Omega$, following the ratio of separation to scatter in Table~\ref{tab:lines}. This is the counterintuitive behavior reported in \cite{r106}, here confirmed with independent data and extended from 1000 to 3200 $\Omega$.

\subsection{Comparison with the optimal limit}
Table~\ref{tab:p} also lists the limit $p_{\max}(100)$ of Eq.~(\ref{eq:pmax}), evaluated with the covariance matrix estimated from the data, and Fig.~\ref{fig:p} shows it as the thin black line. The estimated $\kappa$ is given in Table~\ref{tab:bep}. The asymmetry parameter is $\kappa = 0.055$ at 50 $\Omega$ and decreases slowly to 0.047 at 3200 $\Omega$. The limit is $p_{\max} = 0.708$ at 50 $\Omega$, 0.707 at 200 $\Omega$, 0.701 at 800 $\Omega$ and 0.682 at 3200 $\Omega$, and 0.5 at zero wire resistance. The computer simulation of Eq.~(\ref{eq:pmax}) agrees with the exact formula to three decimals, the Gaussian approximation in Eq.~(\ref{eq:pmax}) agrees with it to four decimals already at $N = 100$, and the measured success probability of the sign rule lies inside its 95\% confidence interval around the limit at every wire resistance. Logistic regression falls 0.003 to 0.006 below the limit, and the nearest-line rule stays within 0.006 of it, slightly above at the smallest wire resistances and below at the largest. The fold-to-fold standard deviation of the cross-validated $p$ is between 0.005 and 0.02. Machine learning therefore approaches the limit to within approximately 0.01 for non-zero wire resistance but cannot exceed it, which is what Eq.~(\ref{eq:lr}) says. The slow decrease of $p$ with increasing wire resistance in Fig.~\ref{fig:p} is the decrease of $\kappa$ in Table~\ref{tab:bep}: the wire resistance increases the difference $\Sigma_{AA} - \Sigma_{BB}$, but it increases $\det\Sigma$ faster, because the wire resistance decouples the noise generators of Alice and Bob \cite{r106}.

\subsection{Dependence on the BEP length}
\begin{figure}[htbp]
\centering\includegraphics[width=4.0in]{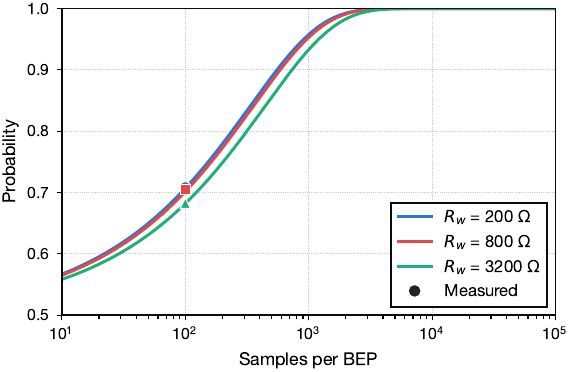}
\caption{The upper limit $p_{\max}$ of Eve's success probability, Eq.~(\ref{eq:pmax}), versus the number of samples per BEP, $N$, for $R_L = 9$ k$\Omega$ and $R_H = 10$ k$\Omega$ at 200, 800 and 3200 $\Omega$ wire resistance, with the asymmetry parameters of Table~\ref{tab:bep}. The markers are the measured sign-rule values at $N = 100$.}
\label{fig:pN}
\end{figure}

Beyond the 100 samples per BEP used in the simulations, Eq.~(\ref{eq:pmax}) also gives the maximum success probability for any other BEP length. Figure~\ref{fig:pN} shows $p_{\max}$ as a function of $N$ for three wire resistances, and Table~\ref{tab:bep} lists the BEP length needed for $p_{\max} = 0.90$, 0.95 and 0.99. At 200 $\Omega$, Eve needs about 550 samples per bit to reach $p = 0.90$, 910 for $p = 0.95$ and 1820 for $p = 0.99$. At 3200 $\Omega$ the values are 730, 1200 and 2410. Since $\kappa$ is fixed by the resistors and the wire resistance, the only way for Eve to increase her success probability at this narrow resistor gap is a longer BEP, and $p_{\max}$ approaches 1 as $\kappa\sqrt{N}$ grows. The BEP length is the number of independent samples in the bit exchange period \cite{r104}. As noted in \cite{r106}, the narrow resistor gap increases Alice's and Bob's own bit error probability at a fixed clock period. Thus the BEP length is the parameter that sets the trade-off between the security against the classifier attack and the speed of the key exchange.

\begin{table}[htbp]
\tbl{Asymmetry parameter $\kappa$ of Eq.~(3) estimated from the data, the upper limit $p_{\max}$ of Eve's success probability, Eq.~(4), for $N = 100$ and $N = 1000$ samples per BEP, and the BEP length $N$ required for $p_{\max} = 0.90$, 0.95 and 0.99.}{
\begin{tabular}{@{}rcccccc@{}}\toprule
$R_w$ ($\Omega$) & $\kappa$ & $p_{\max}(100)$ & $p_{\max}(1000)$ & $N$ for 0.90 & $N$ for 0.95 & $N$ for 0.99 \\ \midrule
0 & 0.0000 & 0.500 & 0.500 & $\infty$ & $\infty$ & $\infty$ \\
50 & 0.0550 & 0.708 & 0.959 & 545 & 897 & 1794 \\
100 & 0.0548 & 0.708 & 0.958 & 548 & 902 & 1803 \\
200 & 0.0546 & 0.707 & 0.958 & 553 & 911 & 1821 \\
400 & 0.0540 & 0.705 & 0.956 & 564 & 929 & 1858 \\
800 & 0.0530 & 0.701 & 0.953 & 587 & 967 & 1933 \\
1600 & 0.0510 & 0.694 & 0.946 & 634 & 1043 & 2086 \\
3200 & 0.0474 & 0.682 & 0.933 & 732 & 1205 & 2409 \\
\bottomrule\end{tabular}}
\label{tab:bep}
\end{table}

\section{Discussion}
The interpretation of these results is straightforward. In the classifier plane, all the information that distinguishes HL from LH is in the sign and size of the difference $x - y$, because the position of a point along the diagonal is set by the noise of the whole loop, which is shared by the two ends through the closed circuit, and is the same for both cases. This is made exact by Eq.~(\ref{eq:lr}): the likelihood ratio depends on $x - y$ alone and grows with it, so the sign rule is the best possible decision, its success probability is the fraction of bits on the correct side of the line $x = y$ in Fig.~\ref{fig:hist}, about 0.71, and no classifier that uses the same BEP can exceed Eq.~(\ref{eq:pmax}). The machine learning classifiers can only approach this limit, and the more complicated ones fall below it, which is consistent with overfitting to finite-sample fluctuations. The AUC of approximately 0.78 indicates that the magnitude of $x - y$ provides useful ranking information even though it does not improve the Bayes-optimal hard decision. The remark in \cite{r104} that machine learning could enhance the attack applies to the well-separated point clouds of strongly asymmetric resistor pairs, where there is no overlap to resolve anyway. With a narrow resistor gap, the overlap is a property of the data and no classifier can remove it.

The decrease of $p$ with increasing wire resistance comes from two competing effects that appear in Table~\ref{tab:lines}: the wire resistance separates the two fitted lines, which helps Eve, but it also decouples the noise generators of Alice and Bob and widens the point clouds, which hurts her. For $R_L = 9$ k$\Omega$ and $R_H = 10$ k$\Omega$ the second effect is slightly stronger over the whole range, so the ratio of separation to scatter, and with it the asymmetry parameter $\kappa$ of Eq.~(\ref{eq:kappa}), decreases slowly and $p$ decreases with it. This is the opposite of the behavior in the classical Bergou-Scheuer-Yariv attack, where the mean-square voltages at the two ends of the wire are simply compared and Eve's signal grows with the square of the wire resistance \cite{r47}. The conclusion of \cite{r106}, that there is a crossover wire resistance at which both attacks give a similar $p$ and which provides a minimax security optimum, is not changed by the present results, because machine learning does not move the curve of the classifier attack.

\begin{figure}[htbp]
\centering\includegraphics[width=4.6in]{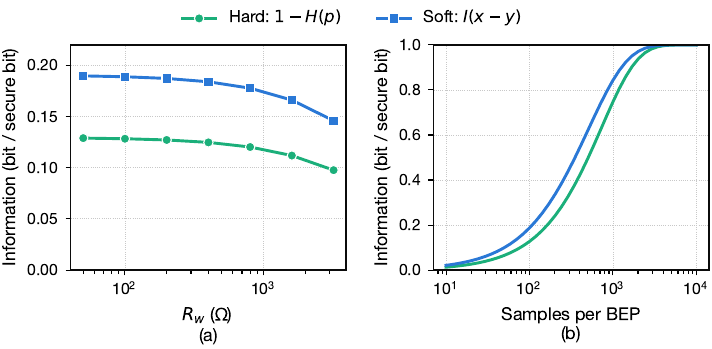}
\caption{Information Eve gains per secure bit. (a) Versus the wire resistance at $N = 100$: the Hard-decision information $1 - H(p)$ is the value usually quoted, while the Soft information, the mutual information between the configuration and the difference $x - y$, is about 1.5 times larger. (b) Both versus the BEP length $N$ at $R_w = 200$ $\Omega$, approaching one bit as $N$ grows.}
\label{fig:mi}
\end{figure}

If Eve reduces each observation to the binary decision produced by the sign rule, then, for equiprobable bits and a symmetric binary decision channel, $p \approx 0.71$ corresponds to $1 - H(0.71) \approx 0.13$ bit per secure bit, where $H$ is the binary entropy function, compared with about one bit for the strongly asymmetric resistor pairs of \cite{r104,r106}. The complete continuous statistic $x - y$, however, also contains information about the confidence of each decision, as reflected by the likelihood ratio and the AUC. Therefore, $1 - H(p)$ characterizes the information associated with the hard-decision channel rather than the total information contained in Eve's continuous observation. This total information can be evaluated directly. Because the log-likelihood ratio of Eq.~(\ref{eq:lr}) depends on the samples only through $x - y$, the difference is a sufficient statistic, and the mutual information between the configuration and $x - y$ is the whole information one BEP gives Eve. For the present data at $N = 100$ it is about 0.19 bit per secure bit (Fig.~\ref{fig:mi}), roughly 1.5 times the hard-decision value $1 - H(p)$, and, like $p_{\max}$, it grows toward one bit as the BEP lengthens. For a BEP of 1000 samples the limit of Eq.~(\ref{eq:pmax}) rises to $p_{\max} = 0.96$, that is, about 0.75 bit per secure bit for the hard-decision channel, so the BEP length is the key parameter for the remaining information. Privacy amplification \cite{r62} can be used to suppress the remaining information available to Eve, with the required compression determined from an appropriate entropy measure of Eve's full observation.

The bit error rate of Alice and Bob was not analyzed in this study. Since the narrow resistor gap increases it at a given BEP length \cite{r106}, it can invalidate the countermeasure.

\section{Conclusions}
We have applied the binary-classifier wire-resistance attack to a KLJN key exchanger with a narrow resistor gap, $R_L = 9$ k$\Omega$ and $R_H = 10$ k$\Omega$, and asked whether supervised machine learning can achieve a higher successful bit-guessing probability than the simple decision rules of \cite{r104,r106}. The results show that it cannot: the nearest-line rule, logistic regression and a support vector machine reproduce the success probability of the sign rule, $p \approx 0.71$ for small wire resistance, while the nearest-neighbor, random forest and gradient boosting classifiers fall below it. The reason is that the distance between the fitted HL and LH lines is about equal to the scatter of the points around them at every wire resistance from 50 to 3200 $\Omega$, so the two point clouds overlap and the sign of the difference between the two mean-square voltages is already the best possible decision. We have shown this exactly: for Gaussian noise the likelihood ratio between the HL and LH cases depends on the samples of a BEP only through this difference, and the success probability of the sign rule has the simple form of Eq.~(\ref{eq:pmax}), which reproduces the measured values and gives the BEP length needed for any target success probability, about 900 to 1200 samples per bit for $p = 0.95$ depending on the wire resistance. The counterintuitive decrease of $p$ with increasing wire resistance, from 0.711 at 50 $\Omega$ to 0.682 at 3200 $\Omega$, is confirmed with independent data and extended beyond the range of \cite{r106}.

These results support resistor-gap narrowing as a physics-level countermeasure against the classifier attack: the hard-decision information associated with $p \approx 0.71$ at $R_L / R_H = 0.9$ is about 0.13 bit per secure bit and does not increase when the eavesdropper replaces the simple sign rule by machine learning. Privacy amplification can be used to suppress the remaining information available to Eve, with the required compression determined from an appropriate entropy measure of Eve's full observation. However, the Bayes-optimal success probability increases with the BEP length according to Eq.~(\ref{eq:pmax}), and consequently the information associated with the hard-decision channel also increases. This is the quantity to be balanced against the bit error rate of Alice and Bob, which is beyond the scope of this study and which may cancel the security gain of the narrow resistor gap. A quantitative treatment of the crossover with the classical wire-resistance attack, and the behavior of $p$ as $R_L$ approaches $R_H$ even more closely, will be the subject of a forthcoming study.

\section*{Author Contributions}
M.Y. developed and executed the time-domain KLJN simulations, generated the classifier datasets, performed the numerical evaluations, prepared figures and tables, and contributed to manuscript drafting. F.A. developed the statistical and machine-learning analysis, including the likelihood-ratio/Bayes-optimality interpretation, classifier-validation methodology, and continuous-information analysis. L.B.K. contributed to the physical KLJN framework and to the manuscript revision. All authors discussed the results and approved the final manuscript.

\renewcommand{\refname}{References}
\begingroup
\setlength{\parskip}{0pt}

\endgroup
\end{document}